\documentclass[preprints,article,accept,moreauthors,pdftex,10pt,a4paper]{mdpi}

\usepackage{amsfonts}
\usepackage{amsmath}
\usepackage{amssymb}
\usepackage{mathtools}
\usepackage{dsfont}
\usepackage{subfig}

\newcommand{\bs}[1]{\boldsymbol{#1}}

\definecolor{darkgreen}{RGB}{50,200,8}

\newcommand{\rev}[1]{\textcolor{black}{#1}}

\graphicspath{ {./} }

\firstpage{1} 
\pubvolume{xx}
\issuenum{1}
\articlenumber{1}
\pubyear{2018}
\copyrightyear{2018}
\externaleditor{}
\history{}

\Title{An information-theoretic approach to self-organisation: Emergence of complex interdependencies in coupled dynamical systems}

\Author{Fernando Rosas $^{1,2}$*, Pedro A.M. Mediano $^3$, Martin Ugarte $^4$ and Henrik J. Jensen $^{1,5}$}

\AuthorNames{Fernando Rosas, Pedro A.M. Mediano, Martin Ugarte and Henrik J. Jensen}

\address{%
$^{1}$ \quad Centre of Complexity Science and Department of Mathematics, Imperial College London, UK\\
$^{2}$ \quad Department of Electrical and Electronic Engineering, Imperial College London, UK\\
$^{3}$ \quad Department of Computing, Imperial College London, UK\\
$^{4}$ \quad CoDE Department,  Universit\'{e} Libre de Bruxelles, Belgium\\
$^{5}$ \quad Institute of Innovative Research, Tokyo Institute of Technology, Japan
}

\corres{Correspondence: f.rosas@imperial.ac.uk; Tel.: +44 (0)20 7589 5111}

\abstract{Self-organisation lies at the core of fundamental but still
unresolved scientific questions, and holds the promise of de-centralised
paradigms crucial for future technological developments. While self-organising
processes have been traditionally explained by the tendency of dynamical
systems to evolve towards specific configurations, or attractors, we see
self-organisation as a consequence of the interdependencies that those
attractors induce. Building on this intuition, in this work we develop a
theoretical framework for understanding and quantifying self-organisation based
on coupled dynamical systems and multivariate information theory. We propose a
metric of global structural strength that identifies when self-organisation
appears, and a multi-layered decomposition that explains the emergent structure
in terms of redundant and synergistic interdependencies. We illustrate our
framework on elementary cellular automata, showing how it can detect and characterise the
emergence of complex structures.
}

\keyword{Self-organisation; multivariate information theory; coupled dynamical systems; partial information decomposition; high-order correlations; multi-layer complexity}

\begin{document}

\vspace{0.5cm}
\section{Introduction}

\subsection{Context}

It is fascinating how some systems acquire organisation spontaneously, evolving
from less to more organised configurations in the absence of centralised
control or an external driver. In a world constricted by the second law of
thermodynamics and driven by ``no free lunch'' principles, self-organisation
phenomena dazzle us by creating structure seemingly out of nowhere. Besides
this aesthetic dimension, self-organisation plays a key role at the core of
out-of-equilibrium statistical physics~\cite{haken1983synergetics},
developmental biology~\cite{camazine2003self}, and
neuroscience~\cite{tognoli2014metastable}. Additionally, self-organisation
serves as inspiration for new paradigms of de-centralised organisation where
order is established spontaneously without relying on an all-knowning architect
or a predefined plan, such as with the Internet of
Things~\cite{ding2013intelligent,athreya2013network} and blockchain
technologies~\cite{macdonald2016blockchains}. In this context,
self-organisation is regarded as an attractive principle for enabling
robustness, adaptability and scalability into the design and managment of
large-scale complex
networks~\cite{prokopenko2013guided,kuze2016controlling,rosas2017technological}.

Originally, the notion of self-organisation was introduced in the field of
cybernetics~\cite{ashby1947principles,foerster1960self}. These seminal ideas
quickly propagated to almost all branches of science, including
physics~\cite{haken1983synergetics,haken2006macroscopic},
biology~\cite{camazine2003self,crommelinck2006self}, computer
science~\cite{heylighen2003meaning,mamei2006case}, language
analysis~\cite{de2000self,steels1998synthesising}, network
management~\cite{prehofer2005self,dressler2008study}, behavioral
analysis~\cite{kugler1980concept,kelso1988self} and
neuroscience~\cite{kelso1997dynamic,friston2010free}. 
Despite
this success, most working definitions of self-organisation still avoid formal
definitions and rely on intuitions following an ``I know when I see it'' logic,
which might eventually prevent further systematic
developments~\cite{shalizi2004quantifying}. Formulating formal definitions of
self-organisation is challenging, partly because self-organisation has been
used in diverse contexts and with different
purposes~\cite{gershenson2012guiding}, and partly due to the fact that the
basic notions of ``self'' and ``organisation'' are already
problematic themselves~\cite{gershenson2003can}.

The absence of an agreed formal definition, combined with the relevance of this
notion for scientific and technological advances, generates a need for further
explorations about the principles of self-organisation.

\subsection{Scope of this Work and Contribution}
\label{sec:1.2}

In the spirit of Reference~\cite{krakauer2014information}, we explore to what extent
an information-theoretic perspective can illuminate the inner workings of
self-organising processes. Due to the connections between information theory
and thermodynamics~\cite{jaynes2003probability,mezard2009information}, our
approach can be seen as an extension of previous works that relate 
self-organisation and statistical physics (see e.g.
\cite{nicolis1977self,heylighen2001science,pulselli2009self}). In previous
research, self-organisation has been associated with a reduction in the
system's
entropy~\cite{nicolis1977self,klimontovich1991turbulent,gershenson2012complexity}
-- in contrast, we argue that entropy reduction alone is not a robust predictor
of self-organisation, and additional metrics are required. 

This work establishes a way of understanding self-organising processes that is
consistent with the Bayesian interpretation of information theory, as described
in Reference~\cite{jaynes2003probability}. One contribution of our approach is to characterise self-organising processes using multivariate information-theoretic tools -- or, put differently, to
provide a more fine-grained description of the underlying phenomena behind
entropy reduction. We propose that self-organising 
processes are driven by spontaneous creation of interdependencies, 
while the reduction of entropy is a mere side effect of this. 
Following this rationale, we propose the
\textit{binding information}~\cite{vijayaraghavan2017anatomy} as a metric of
the strength of the interdependencies in out-of-equilibrium dynamical systems.

Another contribution of our framework is to propose a multi-layered metric of
organisation, which combines quantitative and qualitative aspects. Most
proposed metrics of {organisation in the field of complex systems} try
to map the whole richness of possible structures into a single
dimension~\mbox{\cite{lloyd2001measures}}. In contrast, drawing inspiration
from theoretical
neuroscience~\mbox{\cite{tononi1994measure,friston1995characterising}}, we put
forward a multi-dimensional framework that allows for a finer and more subtle
taxonomy of {
self-organising systems}. Our framework builds on ideas based on
the \textit{Partial Information Decomposition} (PID)
framework~\cite{williams2010nonnegative}, which distinguishes various
information sharing modes in which the binding information is distributed
accross the system. This fundamental distinction overcomes counterintuitive
issues of existent multiscale metrics for structural complexity, such as the
one reported in References~\mbox{\cite{bar2004multiscale,allen2017multiscale}},
including negative information values that do not have operational meaning.

A final contribution of this work is to establish a novel connection between
information theory and dynamical systems. The standard bridge between these two
disciplines includes symbolic dynamics, the Kolmogorov-Sinai entropy, 
R\'enyi dimensions and related
concepts~\cite{beck1995thermodynamics}. In contrast, in this paper we propose
to apply information-theoretic analyses over the statistics induced by
invariant measures over the attractors. In this way, attractors can be seen as
statistical structures that generate interdependencies between the system's
coordinates. This statistical perspective enriches standard
analyses of atractors based on fractal dimensions and other geometrical
concepts.

The rest of this paper is structured as follows. First, Section~\ref{sec:key}
briefly introduces the key ideas of this work. Then, Section~\ref{sec:2}
discusses fundamental aspects of the definition of self-organisation and
coupled dynamical systems. Section~\ref{sec:3} presents the core ideas our
information-theoretic approach, which are then developed quantitavely in
Section~\ref{sec:4}. Our framework is illustrated in Section~\ref{sec:5} with
an application to elementary cellular automata. Finally, Section~\ref{sec:6}
discusses our findings and summarises our main conclusions.


\section{Key Intuitions}
\label{sec:key}

This section introduces the key ideas of our framework in an
intuitive fashion. These ideas are made rigorous in the following sections.

\subsection{The Marble Sculpture Analogy}

The overall configuration of a group of agents can be represented by a
probability distribution over the set of their possible configurations.
Independent agents who are maximally random are characterised by ``flat''
distributions (technically, distributions that satisfy the \textit{maximum
entropy principle}~\cite{jaynes2003probability}). The temporal evolution of the
system then ``shapes'' this distribution, in the same way as a sculptor
shapes a flat piece of marble into a sculpture (c.f. Figure~\ref{fig:entropy_dec}). 
In our view, the shape of the
resulting distribution encodes the key properties that emerge from the temporal
dynamics of the system, and a substantial part of our framework is to provide
tools to measure and describe various types of sculptures. Importantly, just as
the sculptor reveals a figure by removing the superfluous marble that is
covering it, the temporal evolution generates interdependencies not by adding
anything but by reducing the randomness/entropy of the system.

\subsection{Self-Organisation Versus Dissipation}

\begin{figure}[H]	
 \centering
 \includegraphics[width=5in]{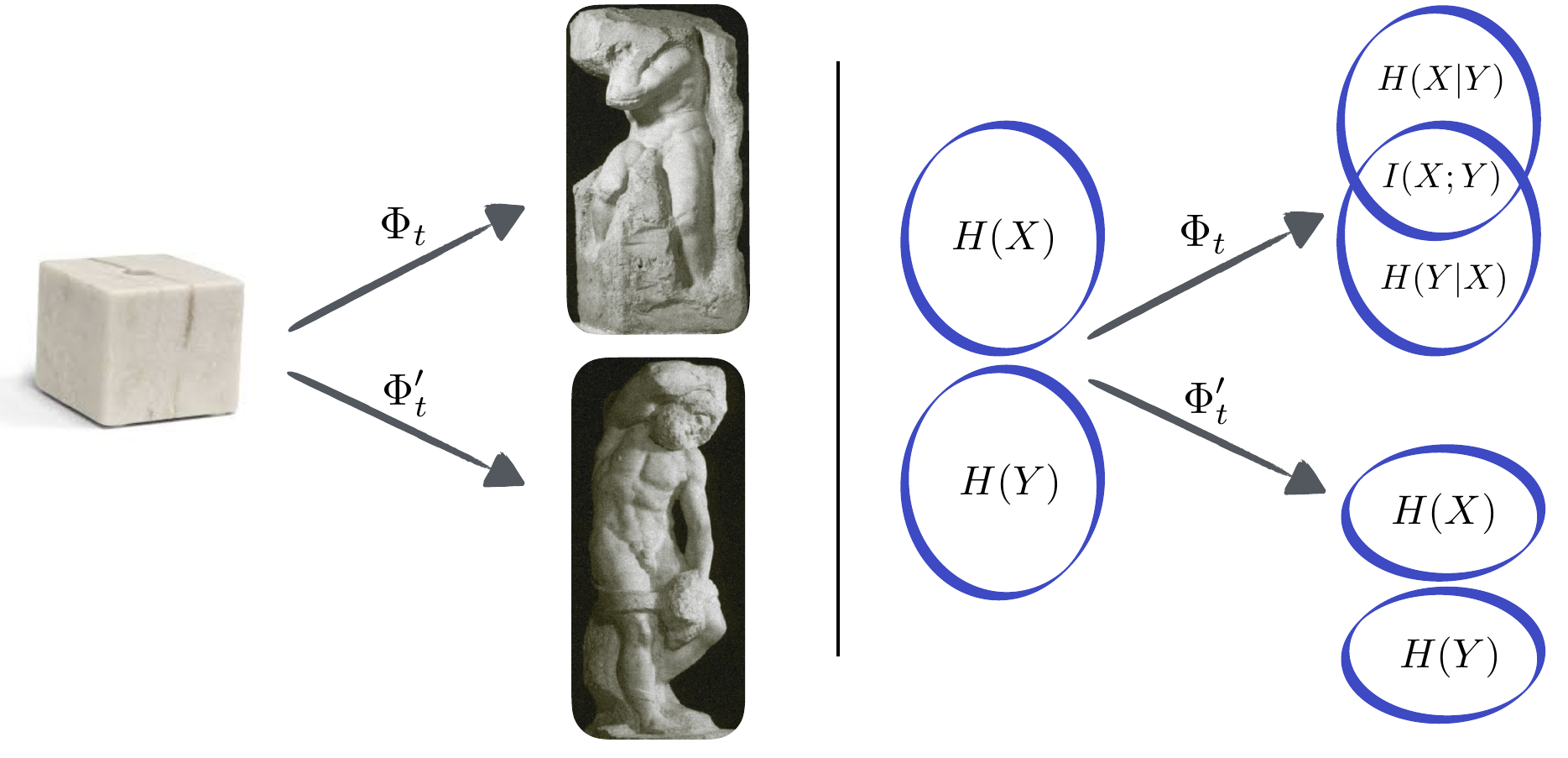}
 \caption{\textit{Left:} Two maps correponding to two dynamical
 systems, denoted by $\Phi$ and $\Phi'$, seen as a sculptor who takes away
 ``superfluous'' entropy/marble to let structures appear from inside. The
 figure shows \emph{The Atlas} and \emph{The Bearded Slave} (circa 1525–30) 
 by Michelangelo Buonarroti,
 who was famous for letting his figures emerge from the marble ``as though
 surfacing from a pool of water''~\cite{vasari1991lives} (pictures taken from
 \url{commons.wikimedia.org}).
 \textit{Right:} Likewise, the joint entropy of two (or more) agents could
 decrease either because they become less random individually ($\Phi'$), or
 because they become correlated ($\Phi$). In this article we provide tools to
 measure how \rev{self-organising} systems shape distributions as entropy is
 reduced -- or marbled is carved out -- from the initial state.}
 \label{fig:entropy_dec}
\end{figure}

Consider two agents with random and uncorrelated initial states, as in the
analogy above. Their joint entropy, which quantifies their collective randomness, 
can be depicted as two circles, the size of each circle being
proportional to how random the corresponding agent is (c.f. Figure~\ref{fig:entropy_dec}). 
The circles are shown disjoint to
reflect the fact that the agents are initially uncorrelated. From this initial
situation, there are two qualitatively different ways in which their joint
entropy can decrease: the state of each agent could become less random in time,
while their independency is preserved; or the agents could become correlated
while their individual randomness is preserved. Although both cases show
overall entropy reduction, one needs to distinguish finer features of the shape
of the resulting distribution to discriminate between genuine self-organisation
in the latter scenario and mere dissipation in the former.

\section{The Goal and Constraints of Self-Organisation}
\label{sec:2}

Which dynamical properties enable agents to self-organise? Beyond superficial
differences, most studies agree that proper self-organisation requires three
fundamental principles to hold:

\begin{enumerate}[leftmargin=*,labelsep=4.9mm]
\item[(i)] \textbf{Global structure:} the system evolves from less to more structured collective configurations.
\item[(ii)] \textbf{Autonomy:} agents evolve in the absence of external guidance.
\item[(iii)] \textbf{Horizontality:} no single agent can determine the evolution of a large number of other agents.
\end{enumerate}

The principles of autonomy and horizontality constitute constraints, 
in the sense that a system that is not autonomous or horizontal cannot be called \textit{self}-organising. Conversely,
the principle of global structure is closer to a goal to be achieved. 
Hence, one could reformulate the above definition of self-organisation
as the following optimisation problem:

\begin{equation}
\begin{aligned}
& \underset{ }{\textit{generate}}
& & \text{Global structure} \\
& \textit{subject to}
& & \text{Autonomy and horizontality}.
\end{aligned}\label{eq:def}
\end{equation}

The following subsections provide a formalisation of these three fundamental principles.

\subsection{Multiple Agents as a Coupled Dynamical System}
\label{sec:22}

An elegant way to formalise these ideas is provided by the literature of
coupled dynamical systems. Loosely speaking, a dynamical system is a process
that evolves in time, such that its present configuration determines its future
evolution following a deterministic rule~\cite{robinson2012introduction}.
Differential equations and finite difference equations are examples of dynamical
systems. Furthermore, a collection of dynamical systems are said to
be coupled if the future state of each process is affected not only by its
own state but also by the state of other processes.

Let us consider a system composed by $N$ parts or subsystems, which we call
``agents'' adopting the terminology from the robotics and multi-agent systems
literature. \rev{However, these agents could correspond to different coordinates of
the spatial movement of a single entity~ \cite{nurzaman2015goal}, or to
sub-systems of heterogenous nature}. The set of
possible states for the $k$-th agent is denoted as $\Omega_k$, and hence the
set of possible configurations of the system is $\Omega \coloneqq \prod_{k=1}^N
\Omega_k$, henceforth called ``phase space.'' The configuration of the system
at time $t \in T\subset [0,\infty)$ is determined by the vector
$\boldsymbol{x}_t = (x_t^1,\dots,x_t^N) \in \Omega$, where $x_t^k\in \Omega_k$
is the corresponding state of the $k$-th agent and $T$ is a collection of time
indices. By assuming that the agents constitute coupled dynamical systems, the
evolution of the group of agents is determined by a collection of maps $\{\phi_{t}^{(h)}\}$ 
with $h \geq 0$, where $\phi_{t}^{(h)}: \Omega
\rightarrow \Omega$ drives the evolution of the system such that
$\boldsymbol{x}_{h+t} = \phi_{t}^{(h)} (\boldsymbol{x}_h)$. Intuitively, $h$
corresponds to an initial time and $t$ is the length of the evolution process.

Please note that the choice of deterministic coupled dynamical systems as the
basis of our framework has been made for simplicity of presentation. The
generalisation of our ideas and methods to stochastic dynamics is
straightforward.

\subsection{Formalising Self-Organisation}
\label{sec:333}

We now discuss aspects of the formalisation of \eqref{eq:def} based on the
language of dynamical systems.

\subsubsection{Autonomy}

Intuitively, we say that a system is autonomous if it has no architect or
``mastermind'' controlling its evolution from the outside. Using the dynamical
systems language introduced above, we can readily define necessary conditions
for the autonomy of a system: we say that a system is autonomous if the
collection of maps $\{\phi_{t}^{(h)}\}$ are time-invariant -- i.e. if
its temporal evolution looks the same independently of the initial time $h$.
Technically, autonomy requires that
\rev{$\phi_{t}^{(h_1)}(\boldsymbol{x}_{h_1})=\phi_{t}^{(h_2)}(\boldsymbol{x}_{h_2})$
for any $h_1,h_2\in T$ and $\boldsymbol{x}_{h_1}=\boldsymbol{x}_{h_2}$}. This
symmetry ensures that there is no organising influence guiding the system from
outside. In the rest of this manuscript time translation symmetry is assumed,
which allows us to disregard the starting time and drop the superscript $(h)$,
using $\phi_{t}$ as a shorthand notation.\footnote{Additionally, autonomous
systems allow simple descriptions. Thanks to the property
$\phi_{t_1}(\phi_{t_2}(\boldsymbol{x})) = \phi_{t_1+ t_2}(\boldsymbol{x})$,
autonomous evolutions in discrete time are characterised by the single maping
$\phi \coloneqq \phi_1$ by noting that $\phi_n = (\phi)^n$, while autonomous
evolutions in continuous time can be characterised by a vector field or a set
of time-invariant differential equations.}

\subsubsection{Horizontality and locality}

Intuitively, horizontality implies a similar restriction within the system itself, in the
sense that no small set of units should control or influence the behaviour of
the rest of the system. 
However, in contrast with the simplicity with which autonomy
can be addressed, the formalisation of horizontality is substantially
more challenging. Our approach here is to take a stronger condition than
horizontality, namely:
\begin{enumerate}[leftmargin=*,labelsep=4.9mm]
\item[(iii-b)] \textbf{Locality:} agents can only interact with a small number of
other agents.
\end{enumerate}

Locality is a sufficient condition for horizontality, since if no agent can
interact with many other agents then the direct influence of each agent is
limited. Conveniently, locality can be elegantly addressed within the framework
of coupled dynamical systems. To do this, let us first introduce the notation
$\phi_{t}^k$ for the $k$-th coordinate of the map $\phi_{t}$, i.e.
$\phi_{t}(\boldsymbol{x}) = (\phi_{t}^1 (\boldsymbol{x}), \dots, \phi_{t}^N
(\boldsymbol{x}))$. Then, one can define the interaction network between agents
as follows: there exists a link from agent $i$ to agent $j$ if
$\phi_{t}^j(\boldsymbol{x})$ is affected by changes in the values of $x_i$, the $i$-th coordinate of
$\boldsymbol{x}$.

These directed networks can be encoded by an $N\times N$ adjacency matrix
$\mathcal{A} = [a_{ij}]$, where $a_{ij} = 1$ if the $i$-th agent is connected
with the $j$-th agent and zero otherwise. Locality is, hence, equivalent to
$\mathcal{A}$ having sparse rows, imposing a fixed bound restricting the number
of non-zero entries in each row.

In the following we assume locality, and leave the formalisation of
horizontality for future work.

\subsubsection{Structure}
\label{sec:23}

One of the biggest challenges in the formalisation of self-organisation is to
address the notion of \textit{structure}. A large portion of the literature
employs this concept without developing a formal definition of it, relying only
on intuitive understanding. Furthermore, authors from different fields
point towards this same intuition using related but different concepts,
including global behaviour, organisation, coordination, or pattern. 

Existing approaches to attempt a formalisation of the notion of structure use
either attractors, or minimal description length and Kolmogorov complexity. These approaches,
and their drawbacks, are discussed in Appendix~\ref{app:alternative_structure}. Our own approach,
which relies in multivariate information theory, is presented in the next Section.


\section{Structure as Multi-Layered Statistical Interdependency}
\label{sec:3}

This section introduces our framework to study emergence of structure in coupled
dynamical systems. The key idea in our approach is to understand
\textit{structure as statistical interdependency} and, hence, to regard
patterns as deviations from statistical independence, i.e. as interdependent random
variables. As argued below, these statistical interdependencies are
best described using tools from multivarate information theory.

Adopting an information-theoretic perspective requires a step of abstraction,
namely to place the analysis not in trajectories but in ensembles, as explained
in Section~\ref{sec:31}. Then, Section~\ref{sec:32} explores the relationship
between the dynamics of the joint Shannon entropy and the increase of
statistical interdependency. This discussion is further developed by
introducing a decomposition of the Shannon entropy in
Sections~\ref{sec:binding_info} and~\ref{sec:34}.

For simplicity of exposition, in the rest of the paper we will focus in the case
of discrete phase space $\Omega$. However, most of our results still hold
for continuous dissipative systems.

\subsection{From Trajectories to Stochastic Processes}
\label{sec:31}

Traditionally, the study of dynamical systems is fundamentally built on how
individual trajectories explore the space of possible system
configurations.\footnote{As a matter of fact, the measure-theoretic objects
that are more studied within dynamical system theory (namely, invariant
measures~\cite{schuster2006deterministic}) are distributions that are derived
from mean values over trajectories.} However, the information-theoretic
perspective works not over trajectories but over ensembles (i.e. probability
distributions). As a matter of fact, associating entropy values to trajectories
is usually problematic, as it involves a number of ad-hoc (and often
unacknowledged) assumptions.\footnote{Technically speaking, a sequence of
symbols in isolation has no Shannon entropy or mutual information because it
involves no uncertainty. The literature usually associates a value of entropy
by considering a stochastic model which most likely generated the sequence.
However, this practice relies on strong assumptions (e.g. ergodicity, or
independence of sucessive symbols), which might not hold in practice. {For the
treatment of this issue by stochastic thermodynamics, see
References~\cite{ao2008emerging,seifert2012stochastic}.}} We make our assumptions
explicit, and develop our analysis on an ensemble of systems initialised with
stochastic initial conditions. The technicalities behind this approach are
developed in the sequel.

Let us consider the case where the initial condition of the system is not a
particular configuration $\boldsymbol{x}_0\in \Omega$, but an ensemble of
configurations described by a probability distribution $\mu_0$. Interestingly,
the map $\phi_t$ not only induces a dynamic on the space of configurations
$\Omega$, but it also induces a dynamic on the space of all probability
distributions over $\Omega$, denoted as $\mathcal{M}(\Omega)$. Consider, as an
example, the discrete distribution $\mu_0 = \sum_{j=1}^\infty c_j
\delta_{\boldsymbol{x}_j}$ where $\delta_{\boldsymbol{x}_j}$ is the {Dirac
delta (or the Kronecker delta if $\Omega$ is discrete)}. For this measure, the
probability of a subset of configurations $O \subset \Omega$ is calculated by
$\mu_0(O) = \sum_{j=1}^\infty c_j \mathds{1}_{\boldsymbol{x}_j}(O)$, where
$\mathds{1}_{\boldsymbol{x}_j}(O) = 1$ if $\boldsymbol{x}_j \in O$ and zero
otherwise. A natural time-evolution of this probability distribution is given
by $\mu_t = \sum_{j=1}^\infty c_j \mathds{1}_{\phi_t (\bs{x}_j)}$. One can
generalise this construction for an arbitrary initial probability distribution
$\mu_0$ by introducing \rev{the Frobenius-Perron operator~\cite{ott2002chaos},
which is} an operator over $\mathcal{M}(\Omega)$ defined as
\begin{equation}\label{eq:flow_measure}
\Phi_t\{ \mu_0 \} (O) \coloneqq \mu_0\big( \phi_t^{-1}(O) \big) = \mu_0\big( \{ \boldsymbol{x} \in \Omega| \phi_t(\boldsymbol{x}) \in O\} \big)
\enspace.
\end{equation}

Note that the collection $\{\Phi_t\{\cdot\}, t\in T\}$ generates a dynamic over
$\mathcal{M}(\Omega)$, and hence constitutes a new dynamical
system.\footnote{Interestingly, there exists a subset of $\mathcal{M}(\Omega)$
that is isomorphic to $\Omega$, namely the set of distributions of the form
$\{\mu_{\boldsymbol{x}} = \mathds{1}_{\boldsymbol{x}} | \boldsymbol{x} \in
\Omega\}$. Therefore, it is consistent to call $\mathcal{M}(\Omega)$ a
\textit{generalised state space}, which corresponds to the notion of ``state''
that is used by quantum mechanics~\cite{breuer2002theory}.}

The set of probability distributions $\{\mu_t = \Phi_t\{ \mu_0 \}, t\in
T\}$ induces a corresponding multivariate stochastic process $\bs{X}_t =
(X_t^1,\dots,X_t^N) = \phi_t(\bs{X}_0)$, which follows a joint probability
distribution $p_{\boldsymbol{X}_t} = \mu_t$ (for the complete statistics of $\bs{X}_t$ 
and technical details of this correspondence, see Appendix~\ref{ap:technical}). Note that the properties of
this stochastic process are completely determined by the initial distribution
$\mu_0$ and the map $\phi_t$. Each sub-process $X_t^k$ describes the
uncertainty related to the state of the agent $k$ at time $t$, the statistics
of which are found by marginalising the joint statistics of
$p_{\boldsymbol{X}_t}$. The aim of the next subsections is to
explore the statistiscal interdependencies that can exist among these
sub-processes.

\subsection{Information Dynamics}
\label{sec:32}

The joint Shannon entropy of the system at time $t$, given by
$H(\boldsymbol{X}_t) \coloneqq -\sum_{\boldsymbol{x}\in\Omega}
p_{\boldsymbol{X}_t}(\boldsymbol{x}) \log
p_{\boldsymbol{X}_t}(\boldsymbol{x})$, corresponds to the information required
to resolve the uncertainty about the state of the system at time $t$ (see
Appendix~\ref{ap:entropy}). The uncertainty reflected by this entropy has two
sources~\cite{schreiber1995noise}. One source is stochasticity in the initial
condition, i.e. when the initial configuration of the system at time $t=0$ is
not fully determined, but only prescribed by a probability distribution. The
second source of uncertainty are stochastic dynamics (also known as ``dynamical
noise''), i.e. when the time evolution could make the system potentially
transit from a single starting configuration to two or more different future
configurations. Dynamical systems have deterministic transitions, and hence
only exhibit the first type of uncertainty.

When considering discrete phase spaces, the deterministic dynamics guarantee
that the uncertainty due to random initial conditions cannot increase; it can
only decrease or be conserved. As a simple example, let us consider a dynamical
system with a single point attractor: even if one does not know where a
trajectory starts, one knows that the trajectory ends in the attracting point.
In this case, any information encoded in the initial condition is vanished by
the dynamics, as one cannot find out where trajectories are coming from. We
call this phenomenon ``information dissipation,'' which mathematically can be
stated as:
\begin{equation}\label{eq:information_dissipation}
H(\boldsymbol{X}_t)  \geq H(\boldsymbol{X}_{t+h}) \qquad \text{for all } \: h> 0
\enspace.
\end{equation}

Due to the deterministic nature of the time evolution,
Equation~\eqref{eq:information_dissipation} is guaranteed by the data-processing
inequality~\cite{cover2012elements}. This decrease in entropy is not in
contradiction with the second law of thermodynamics, as these systems are
generally open and connected to an
environment~\cite{schulman1997time,breuer2002theory}. The equality in
Equation~\eqref{eq:information_dissipation} is attained by closed systems, being
this a direct consequence of the well-known Liouville
theorem~\cite{martynov1995liouville}.

Information dissipation is directly related with the action of attractors. For
a given attractor $A$, its basin of attraction $B(A)$ is the largest subset of
$\Omega$ such that $\lim_{t\rightarrow\infty}\phi_t(\boldsymbol{x}) = A$ for
all $\boldsymbol{x}\in B(A)$. Intuitively, any trajectory starting in $B(A)$
asymptotically runs into $A$. 
Similarly, the evolution of an initial distribution $\mu_0$ supported on $B(A)$
eventually ends up being supported almost only on $A$ when $t$ is large enough;
correspondingly, its Shannon entropy tends to decrease
due to the reduced portion of the phase space where the system is confined to
dwell. 
As such, information dissipation (i.e. entropy decreasing due to the action of 
attractors) is a necessary condition for self-organisation.

It is tempting to postulate entropy reduction as a strong indicator of
self-organisation, based on a loose interpretation of entropy as a metric of
disorder. However, the relationship between entropy and disorder is problematic, as
disorder has different meanings in various contexts and there exists no single
widely accepted definition for it. Moreover, entropy reduction is not a
sufficient condition for self-organisation~\cite{shalizi2004quantifying}. For
example, consider a group of uncopled damped oscillators initialised with
random initial positions and velocities. This system evolves towards the
resting state where all velocities are zero, which is the only point attractor
of the system -- thereby reducing its entropy to zero. However, one would not
want to call this evolution as one that is promoting self-organisation, as the
agents are never engaged in any interaction.

\subsection{Binding and Residual Information}
\label{sec:binding_info}

A key idea that emerges from the previous discussion is to relate organisation
with agent interdependency. Following this rationale, we propose that
self-organisation is related to the increase of interdependency between the
agents due to the dynamics. To formalise this intuition, we explore a
decomposition of the total entropy in two parts: one that quantifies
interaction and one that measures uncorrelated variability.

To introduce the decomposition, let us first consider the following identity:
\begin{equation}
H(X_t^j) = I(X_t^j \,; \, \boldsymbol{X}_t^{-j} ) + H(X_t^j | \boldsymbol{X}_t^{-j} )
\enspace,
\end{equation}
where we are using the shorthand notation $\boldsymbol{X}_t^{-j} =
(X_t^1,\dots,X_t^{j-1}, X_t^{j+1},\dots,X_t^N)$, and $I(\cdot \, ; \, \cdot)$ is the
standard Shannon mutual information. This equality states that the entropy of
the state of the $j$-th agent, as quantified by $H(X_t^j)$, can be decomposed
into a part that is shared with the other agents, $I(X_t^j;
\boldsymbol{X}_t^{-j} )$, and a part that is not, $H(X_t^j |
\boldsymbol{X}_t^{-j} )$. This intuition is made rigurous by the
Slepian-Wolf coding scheme~\cite{slepian1973noiseless, el2011network}, which
shows that $I(X_t^j; \boldsymbol{X}_t^{-j} )$ corresponds to information about
the $j$-th agent that can be retrieved by measuring other agents, while
$H(X_t^j | \boldsymbol{X}_t^{-j} )$ is information that can only be retrieved
by measuring the $j$-th agent.

Following the above rationale, the total ``non-shared information'' in the
system is nothing more than the sum of the non-shared information of every
agent, and corresponds to the \textit{residual
entropy}~\cite{vijayaraghavan2017anatomy}:
\begin{equation}
R(\boldsymbol{X}_t) \coloneqq \sum_{j=1}^N H(X_t^j | \boldsymbol{X}_t^{-j})
\enspace.
\end{equation}

One can verify that the agents are statistically independent at time $t$ if and
only if $H(\boldsymbol{X}_t) = R(\boldsymbol{X}_t)$. The complement of the
residual entropy corresponds to the \textit{binding
information}~\cite{te1978nonnegative}, which quantifies the part of the joint
Shannon entropy that is shared among two or more agents. This can be computed
as
\begin{equation}
B(\boldsymbol{X}_t) \coloneqq H(\boldsymbol{X}_t) - R(\boldsymbol{X}_t) = H(\boldsymbol{X}_t) - \sum_{j=1}^N H(X_t^j | \boldsymbol{X}_t^{-j})
\enspace.
\end{equation}

Note that the above formula corresponds to a multivariate generalisation of the
information-theoretic identity $I(X;Y) = H(X,Y) - H(X|Y) - H(Y|X)$, which
captures linear and non-linear dependecies that might exist between two or more
agents. As such, the binding information is one of several multivariate
generalisations of the mutual information, and is the only one known to enable
a non-negative decomposition of the joint
entropy~\cite{rosas2016understanding}.

In summary, the binding information provides a natural metric of the strength
of the statistical interdependencies within a system. {In fact, this metric
is consistent with the intuition that a faithful metric of organisational
richness should be small for systems with maximal or minimal joint entropy (see
Reference~\cite{feldman1998measures} and references therein). On the one hand,
maximal entropy takes place when agents are independent, which implies that
$H(\boldsymbol{X}_t) = R(\boldsymbol{X}_t)$ and hence $B(\boldsymbol{X}_t)=0$
due to the lack of interaction. On the other hand, entropy is minimised in
systems that exhibit no diversity, which limits their binding information due
to the fact that $B(\boldsymbol{X}_t) \leq H(\boldsymbol{X}_t) = 0$.}

Interestingly, although the deterministic nature of deterministic dynamical
systems constrains $H(\boldsymbol{X}_t)$ to be non-increasing, both
$B(\boldsymbol{X}_t)$ and $R(\boldsymbol{X}_t)$ can increase or decrease. In
contrast with the entropy, an increase in binding information is an unequivocal
sign that statistical structures are being generated within the system by its
temporal evolution.

\subsection{The Anatomy of the Interdependencies}
\label{sec:34}

Although the binding information provides an attractive information-theoretic
metric of organisation strength, a one-dimensional description is not rich
enough to describe the range of phenomena observed in self-organising agents.
To obtain a more detailed picture we use the Partial Information Decomposition
(PID) framework, which allows us to develop a finer decomposition of the
binding information and distinguish between different modes of information
sharing. Originally, PID was introduced to study various aspects of
information-theoretic inference, which consider a target variable predicted
using the information provided by a number of information sources (see
  References~\cite{williams2010nonnegative,olbrich2015information,barrett2015exploration,lizier2018information}
and references therein). A key intuition introduced by these works is to
distinguish between various information modes: in particular, \textit{redundant
information} corresponds to information about the target variable that can be
retrieved from more than one source, and \textit{synergistic information}
corresponds to information that becomes available only when two or more sources
are accessed simultaneously.

Traditional PID approaches divide the variables between target and sources,
having each of them a very different role in the framework. Nevertheless, it is
possible to propose symmetric decompositions of the joint Shannon entropy using
PID principles that avoid these dialectic
labellings~\cite{Rosas2015benelux,rosas2016understanding,ince2017partial}. In
this case, the total information encoded in the system's configuration is
decomposed in redundant, unique and synergistic components. Redundancy takes
place when measuring a single agent allows the observer to predict the state of
other agents. Synergy corresponds to high-order statistical effects that can
constrain groups of variables without imposing low-order restrictions. This
idea of synergistic information is a generalisation of the well-known fact that
random variables can be pairwise independent while being jointly
interdependent. The relationship between synergistic information and high-order
correlations in the context of statistical physics has been explored in
References~\cite{rosas2016understanding,olbrich2015information}.

The work reported in \cite{rosas2016understanding} describes a decomposition
for the binding information for the case of systems of $N=3$ agents. In the
sequel, we extend these ideas postulating three formal decompositions of the
binding information for larger system sizes. Please note that our approach here
is not to establish precise formulas for computing the value of the components
of these decompositions for arbitrary underlying probability distributions.
Instead, in Section~\ref{sec:41} we provide universal upper and lower bounds
for these components, which need to be satisfied irrespective of the chosen
functional form. Moreover, these bounds can be used in some cases to determine
exact values of the decomposition's components, as illustrated in
Section~\ref{sec:5}.

\subsubsection{{Decomposition by extension of sharing}}
\label{sec:341}

Since the binding information is the information shared by two or more agents,
it is natural to discriminate exactly how many agents are involved in the
sharing. Following this rationale, we propose the following decomposition:
\begin{equation}\label{eq:dec_dtc}
B(\boldsymbol{X}_t) = \sum_{n=2}^N b^{n}(\boldsymbol{X}_t)
\enspace,
\end{equation}
where $b^{n}(\boldsymbol{X}_t)$ measures the portion of the binding information
that is shared among exactly $n$ agents. The index $n$ refers to the number of
agents that are linked by the corresponding relationship. Therefore,
$b^{n}(\boldsymbol{X}_t)$ quantifies the strength of interdependencies that
link groups of $n$ agents.

To illustrate these ideas, let us explore some simple examples where
this decomposition can be computed directly from our desiderata.

\begin{Example}\label{ex1}
Consider two independent Bernoulli random variables $U$ and $V$ with parameter $p=0.5$ (i.e
$H(U)=H(V)=1$). Then,
\begin{itemize}[leftmargin=*,labelsep=4.9mm]
\item[(i)] If $(X_t^1, X_t^2, X_t^3) = (U,U,U) $, then $R(\boldsymbol{X}_t) = 0$ and $B(\boldsymbol{X}_t) = H(U)$.
Furthermore, because of the triple identity $b^{3}(\boldsymbol{X}_t)=H(U)$, and hence $b^{2}(\boldsymbol{X}_t)=0$.
\item[(ii)] If $(X_t^1, X_t^2, X_t^3) = (U,U,V) $, then $R(\boldsymbol{X}_t) = H(V)$ and $B(\boldsymbol{X}_t) = H(U)$.
In this case, $b^{2}(\boldsymbol{X}_t)=H(U)$ and hence $b^{3}(\boldsymbol{X}_t)=0$.
\item[(iii)] If $(X_t^1, X_t^2, X_t^3) = (U,V,U \texttt{(xor)} V) $, then $R(\boldsymbol{X}_t) = 0$ and $B(\boldsymbol{X}_t) = H(U)+H(V)$.
Furthermore, due to the triple interdepedency $b^{3}(\boldsymbol{X}_t)=H(U)+H(V)$, and hence $b^{2}(\boldsymbol{X}_t)=0$.
\end{itemize}
\end{Example}

\subsubsection{Decomposition by sharing modes}
\label{sec:342}

Following Reference~\cite{rosas2016understanding}, we distinguish between redundant
and synergistic information sharing modes. Redundancy, in this context, refers
to information that is disclosed as soon as any of the agents who
participate in the sharing are measured. Put differently, if agents are engaged
in a redundant information sharing mode then measuring one of them allows the
observer to make inferences on the states of the others. Conversely,
synergistic information sharing takes place when accessing the state of one
agent is not enough to obtain predictive power, i.e. to infer the state of the
other agents involved. The key element is then how many agents need to be
measured in order to obtain information about the other agents. Synergistic
relationships require two or more, implying high-order statistical effects.

Based on these ideas, we postulate the following decomposition:
\begin{equation}\label{eq:dec1}
b^n(\boldsymbol{X}_t) = \sum_{i=1}^{n-1} I_i^n(\boldsymbol{X}_t) ~ ,
\end{equation}
where $I_i^n(\boldsymbol{X}_t)$ denotes information that is shared between $n$
agents, and becomes fully available after accessing $i<n$ of the agents
involved in the sharing. In other words, $i$ is the smallest number of agents
that enables the use of the information that corresponds to
$I_i^n(\boldsymbol{X}_t)$ for predicting the state of the remaining $n-i$
agents. Note the use of upperscripts and lowerscripts differentiate between
group sizes and order of the sharing mode.
This decomposition introduces a range of
$(i,n)$-interdependencies, where $n$ is the extension of the interdependency
(how many agents are involved) while $i$ is the ``degree of synergy.'' With
this notation, redundancies correspond to $i=1$-interdependencies, while
$I_i^n(t)$ for $i\geq 2$ are synergies of order $i$.

Based on these ideas, another way of decomposing the binding information is by
focusing on the possible \textit{information sharing modes}, i.e. ways in which
information can be shared among the agents according to $i$. By combining
Equations~\eqref{eq:dec_dtc} and \eqref{eq:dec1}, one can then present the following
decomposition:
\begin{equation}
B(\boldsymbol{X}_t) = 
\sum_{n=2}^N \sum_{i=1}^{n-1}  I_i^n(\boldsymbol{X}_t) = \sum_{i=1}^{N-1} m_i(\boldsymbol{X}_t) ~ ,\label{eq:dec_modes}
\end{equation}
where $m_i(\boldsymbol{X}_t) \coloneqq \sum_{n=i+1}^N I_i^n(\boldsymbol{X}_t)$
corresponds to information sharing modes that are fully accessed when measuring
sets of $i$ agents. In particular, $m_1(\boldsymbol{X}_t)$ collects all the
``redundancies'' of the system, i.e. sharing modes that are fully accessed by
measuring only one of the agents involved in the sharing. Correspondingly, the
terms $m_i(\boldsymbol{X}_t)$ for $i\geq 2$ convey the strength of synergies
and high-order effects.

To contrast these ideas with the previous decomposition, we study the same
scenarios from Example~\ref{ex1} under this new perspective.

\begin{Example}\label{ex2}
Consider $U$ and $V$ as defined in Example~\ref{ex1}. Then,
\begin{itemize}[leftmargin=*,labelsep=4.9mm]
\item[(i)] If $(X_t^1, X_t^2, X_t^3) = (U,U,U) $, then $m_1(\boldsymbol{X}_t) =H(U)$,
as the information contained in any variable allows to predict the others, while $m_2(\boldsymbol{X}_t)=0$.
\item[(ii)] If $(X_t^1, X_t^2, X_t^3) = (U,U,V) $, then similarly as above $m_1(\boldsymbol{X}_t) =H(U)$ and $m_2(\boldsymbol{X}_t)=0$. 
Both cases are redundancies (same $i$) of disimilar extension (different $n$).
\item[(iii)] If $(X_t^1, X_t^2, X_t^3) = (U,V,U \texttt{(xor)} V) $, then measuring one agent does not allow any predictions over the others, while
by measuring two agents one can predict the third one.\footnote{For a discussion on the statistical properties of the $\texttt{xor}$, 
please see Reference~\cite{rosas2016understanding}.} This implies that
$m_2(\boldsymbol{X}_t) =H(U)+H(V)$, and hence $m_1(\boldsymbol{X}_t)=0$.
\end{itemize}
\end{Example}


\section{A Quantitative Method to Study Time-Evolving Organisation}
\label{sec:4}

In this section we leverage the ideas discussed in Section~\ref{sec:3} to
develop a method to conduct a quantitative and qualitative analysis of
self-organisation in dynamical systems. The goal of this method is twofold: to
detect when self-organisation is taking place, and to characterise it as
redundancy- or synergy-dominated. For this, Section~\ref{sec:41} first develops
upper and lower bounds for the terms of the decompositions of the binding
information presented in Section~\ref{sec:34}. Then, Section~\ref{sec:method}
outlines a protocol of four steps that can be applied in practical scenarios.

\subsection{Bounds for the Information Decompositions}
\label{sec:41}

\subsubsection{Upper bounds for the decomposition by extension}

Let us define $\boldsymbol{\alpha}_L=(\alpha_1,\dots,\alpha_L)$ to be a vector
of $L$ integer indices with $1\leq \alpha_1 < \alpha_2 <\dots< \alpha_L \leq N$,
and $B(\boldsymbol{X}_t^{\boldsymbol{\alpha}_L})$ to be the binding information 
of the agents that correspond to those indices at \linebreak time $t$, i.e.
\begin{equation}
B(\boldsymbol{X}_t^{\boldsymbol{\alpha}_L})   = H(\boldsymbol{X}_t^{\boldsymbol{\alpha}_L}) - \sum_{j=1}^L H(X_t^{\alpha_j} | X_t^{\alpha_1},\dots, X_t^{\alpha_{j-1}}, X_t^{\alpha_{j+1}},\dots, X_t^{\alpha_L} )
\enspace,
\end{equation}
where $\boldsymbol{X}_t^{\boldsymbol{\alpha}_L} = (X_t^{\alpha_1},\dots, X_t^{\alpha_L})$. 
Also, let us denote as $\mathcal{I}_L$ the set of all index
vectors $\boldsymbol{\alpha}_L$ of length $L$, which correspond to the
possible subsets of $L$ agents with cardinality $| \mathcal{I}_L | =
\binom{N}{L}$.

Recall that $b^n(\boldsymbol{X}_t)$ corresponds to information that is shared
exactly by $n$ agents, and hence $\sum_{n=2}^{L}b^n(\boldsymbol{X}_t)$ is the
information shared by $L$ or less agents. As
$B(\boldsymbol{X}_t^{\boldsymbol{\alpha}_L})$ corresponds to the information
shared between agents $\alpha_1,\dots,\alpha_L$
it is clear that for any $L\in\{2,\dots,N\}$ the following bounds hold:

\begin{equation}\label{eq:bounds1}
\sum_{n=2}^{L}b^n(\boldsymbol{X}_t) 
\leq \sum_{\boldsymbol{\alpha}_L \in \mathcal{I}_L} B(\boldsymbol{X}_t^{\boldsymbol{\alpha}_L}) 
\leq \binom{N}{L}  \max_{\boldsymbol{\alpha}_L \in \mathcal{I}_L}  B(\boldsymbol{X}_t^{\boldsymbol{\alpha}_L})
\enspace.
\end{equation}

Although these bounds might not be tight, Equation \eqref{eq:bounds1} suggests that
$\max_{\boldsymbol{\alpha}_L \in \mathcal{I}_L} B_{\boldsymbol{\alpha}_L}(t)$
can be useful for sizing the value of $\sum_{n=2}^L b^n(t)$. In particular,
if $\max_{\boldsymbol{\alpha}_L \in \mathcal{I}_L} B_{\boldsymbol{\alpha}_L}(t)
= 0$ then $b^n(\boldsymbol{X}_t) = 0$ for all $n=2,\dots,L$, which due to Equation~\eqref{eq:dec_dtc} 
would imply that $B(\boldsymbol{X}_t) =\sum_{n=L+1}^Nb^n(\boldsymbol{X}_t) $.

{These bounds are illustrated in the following example.
\begin{Example}
Consider $U$ and $V$ as defined in Example~\ref{ex1}. Let us focus in $L=2$, and note that for this case $\mathcal{I}_2=\{ \{1,2\},\{1,3\},\{2,3\}\}$, and hence
\begin{equation}
\max_{\boldsymbol{\alpha}_2 \in \mathcal{I}_2} B(\boldsymbol{X}_t^{\boldsymbol{\alpha}_2}) = \max_{i,j\in\{1,2,3\}} I(X_t^i;X_t^j)
\qquad
\text{and}
\qquad
\sum_{\boldsymbol{\alpha}_2 \in \mathcal{I}_2} B(\boldsymbol{X}_t^{\boldsymbol{\alpha}_2}) = \sum_{i=1}^3 \sum_{j=i+1}^3 I(X_t^i;X_t^j)
\enspace.
\end{equation}

Using this, it is direct to find that:

\begin{itemize}[leftmargin=*,labelsep=4.9mm]
\item[(i)] If $(X_t^1, X_t^2, X_t^3) = (U,U,U) $, then $\binom{3}{2} \max_{\boldsymbol{\alpha}_2 \in \mathcal{I}_2} B(\boldsymbol{X}_t^{\boldsymbol{\alpha}_2})=\sum_{\boldsymbol{\alpha}_2 \in \mathcal{I}_2} B(\boldsymbol{X}_t^{\boldsymbol{\alpha}_2}) = 3 H(U)$, and hence Equation~\eqref{eq:bounds1} shows that  $b^2(\boldsymbol{X}_t)\leq 3 H(U)$. This bound is not tight, as $b^2(\boldsymbol{X}_t^3) =0$ (c.f. Example~\ref{ex1}). Also, note that for $L=3$ one finds that $ \max_{\boldsymbol{\alpha}_3 \in \mathcal{I}_3} B(\boldsymbol{X}_t^{\boldsymbol{\alpha}_3}) = B(\boldsymbol{X}_t) = H(U)$, showing that the bounds don't need to be monotonic on $L$.
\item[(ii)] If $(X_t^1, X_t^2, X_t^3) = (U,U,V) $, then $\max_{\boldsymbol{\alpha}_2 \in \mathcal{I}_2} B(\boldsymbol{X}_t^{\boldsymbol{\alpha}_2}) = \sum_{\boldsymbol{\alpha}_2 \in \mathcal{I}_2} B(\boldsymbol{X}_t^{\boldsymbol{\alpha}_2}) = H(U)$. This bound is tight, as $b^2(\boldsymbol{X}_t) = H(U)$ (c.f. Example~\ref{ex1}).
\item[(iii)] If $(X_t^1, X_t^2, X_t^3) = (U,V,U \texttt{(xor)} V) $, then $\max_{\boldsymbol{\alpha}_2 \in \mathcal{I}_2} B(\boldsymbol{X}_t^{\boldsymbol{\alpha}_2})=0$, and hence the bounds determine that $b^2(\boldsymbol{X}_t)=0$.
\end{itemize}
\end{Example}
}

\subsubsection{Upper and lower bounds for the decomposition by sharing modes}

Let us recall that $m_i(\boldsymbol{X}_t)$ accounts for the information about
other agents that is obtained when measuring groups of $i$ agents, but not
less. Similarly, $\sum_{i=1}^L m_i(\boldsymbol{X}_t)$ is the predictability
about other agents that is obtained when accessing $L$ or less agents. 
Therefore, one can provide the following bounds valid
for any $L\in\{1,\dots,N-1\}$:
\begin{equation}\label{bounds2}
\psi_L(t) \leq \sum_{i=1}^L m_{i}(\boldsymbol{X}_t) 
\leq \sum_{j=1}^N \sum_{\substack{\boldsymbol{\alpha}_L \in \mathcal{I}_L \\ \alpha_i \neq j}} I(\boldsymbol{X}^{\boldsymbol{\alpha}_L}_t;X^j_t) 
\leq N \binom{N-1}{L} \psi_L(t)
\enspace,
\end{equation}
where we have used the shorthand notation
\begin{align}
  \psi_L(t) \coloneqq \max_{j\in\{1,\dots,N\}} \max_{\substack{\boldsymbol{\alpha}_L \in \mathcal{I}_L\\ \alpha_i \neq j}} I(\boldsymbol{X}^{\boldsymbol{\alpha}_L}_t;X^j_t) ~ .
\end{align}
As in Equation~\eqref{eq:bounds1}, this shows that $\psi_L(t)$ can be used as a
proxy for estimating the relevance of $\sum_{i=1}^L m_{i}(\boldsymbol{X}_t)$.
In particular, if $\psi_L(t) = 0$ then $\sum_{i=1}^L m_{i}(\boldsymbol{X}_t) =
0$. Therefore, by using Equation~\eqref{eq:dec_modes}, if $\psi_L(t) = 0$ then all
the binding information is composed by synergies of order $L+1$ or more.

This discussion suggests that the properties of $\psi_L(t)$ can reveal the
distribution of sharing modes across the system. First, note that $\psi_L(t)$
is a non-decreasing function of $L$: information (in the Shannon sense)
``never hurts,'' and hence having larger groups of agents for making
predictions cannot reduce predictive power. Secondly, in most scenarios
$\psi_L(t)$ is concave: the additional perdictability obtained by including one
more agent usually shows diminishing returns as $L$ grows. In effect, the most
informative agents are normally selected first, and hence for large values of
$L$ one can just add agents with weak informative power, which can also be
redundant with the agents already considered. Accordingly, scenarios where
$\psi_L(t)$ as function of $L$ is concave are called
\textit{redundancy-dominated}. In contrast, scenarios in which $\psi_L(t)$ is
convex are called \textit{synergy-dominated}. Intuitively, in synergy-dominated
scenarios agents might be uninformative by themselves, but become informative
when grouped together. Therefore, a convex $\psi_L(t)$ is a sign of a
synergistic system, one that has larger predictability gains when $L$ grows.

{These ideas and bounds are illustrated in the following example.
\begin{Example}
Consider again $U$ and $V$ as defined in Example~\ref{ex1}. Focusing in $L=1$, one finds that $\psi_1(t) = \max_{i,j\in\{1,\dots,N\}} I(X_t^i;X_t^j)$. 
Therefore, one can find that:
\begin{itemize}[leftmargin=*,labelsep=4.9mm]
  \item[(i)] If $(X_t^1, X_t^2, X_t^3) = (U,U,U) $, then $\psi_1(t)=H(U)$. Therefore, the bounds in Equation~\eqref{bounds2} show that  $H(U)\leq m_1(\boldsymbol{X}_t)\leq 3 H(U)$.
\item[(ii)] If $(X_t^1, X_t^2, X_t^3) = (U,U,V) $, then again $\psi_1(t) = H(U)$, hence the bounds are the same as above.
\item[(iii)] If $(X_t^1, X_t^2, X_t^3) = (U,V,U \texttt{(xor)} V) $ then $\psi_1(t)=0$, which in turn guarantees that $m_1(\boldsymbol{X}_t)=0$.
\end{itemize}
By noting that $\psi_2(t) = \max\{ I(X_t^1;X_t^2X_t^3), I(X_t^2;X_t^1X_t^3),
I(X_t^3;X_t^1X_t^2)\}$, a direct calculation shows that $\psi_2(t) = H(U)$ for the
three above cases. By considering $\psi_0(t) \coloneqq 0$, one finds that cases
(i) and (ii) are redundancy-dominated, while case (iii) is
synergy-dominated.
\end{Example}
}

\subsection{Protocol to Analyse Self-Organisation in Dynamical Systems}
\label{sec:method}

Wrapping up these results, we propose the following definitions for
self-organisation. Note that these are aimed at quantifying organisation,
while the constraints of ``self'' are guaranteed by restricting to autonomous
maps with sparse interaction matrices (see Section~\ref{sec:333}).

\begin{Definition}\label{def1}
Consider a coupled dynamical system with autonomous evolution and a
bounded number of non-zero elements per row in its interaction matrix. Then,
the system is self-organising if $B(\boldsymbol{X}_t)$ is an increasing
function of $t$. Moreover, the value of $B(\boldsymbol{X}_t)$ is used as a
metric of organisation strength.
\end{Definition}

\begin{Definition}\label{def2}
A self-organising process is said to be
\textit{synergy-dominated} if \rev{$\lim_{t\rightarrow \infty}\psi_L(t)$} is convex as function of $L$.
If \rev{$\lim_{t\rightarrow \infty}\psi_L(t)$} is concave, the process is said to be
\textit{redundancy-dominated}.
\end{Definition} 

\rev{Note that for certain processes $\lim_{t\rightarrow \infty}\psi_L(t)$ can
exhibit a combination of convex and concave segments, which suggests the
coexistence of redundant and synergistic structures at different scales. An
example of this is discussed in Section~\ref{sec:mixed_profile}.}

Following these definitions, we propose the following protocol for analysing a given
dynamical system. The steps are:

\begin{itemize}[leftmargin=*,labelsep=4.9mm]

\item[(0)] Check that the maps satisfy autonomy and locality
(Section~\ref{sec:333}).

\item[(1)] Consider a random initial condition given by a uniform distribution
over the phase space, $\mu_0$, and use it to drive the coupled dynamical
system. This involves initialising the system in the least biased initial
configuration, i.e. with maximally random and independent agents.

\item[(2)] Compute the evolution of the probability distribution given by
$\mu_t = \Phi_t\{\mu_0\}$. This can be done directly using the
map, a master equation~\cite{van1992stochastic}, or in the case of a
finite phase space by computing numerically all the trajectories.

\item[(3)] Compute the joint Shannon entropy $H(\boldsymbol{X}_t)$, the
residual information $R(\boldsymbol{X}_t)$, and the binding information
$B(\boldsymbol{X}_t)$ as a function of $t$.

\item[(4)] For values of $t_0$ at which $B(\boldsymbol{X}_{t_0}) > 0 $, compute
$\psi_L(t_0)$ for $L=1,\dots,N$.

\end{itemize}

Note that by considering a flat initial condition in step (1), one ensures that
the system initially has no correlations, i.e. $B(\boldsymbol{X}_0) = 0$.
Therefore, if one finds that $B(\boldsymbol{X}_t) >0$ for some $t>0$, one can
be sure that these interdependencies were entirely created by the dynamics of
the system. Also, while step (3) clarifies if self-organisation is taking place
following Definition~\ref{def1}, (i.e. by checking if $B(\boldsymbol{X}_t)>0$
for some $t>0$), step (4) discriminates between redundant and synergistic
organisation structures according to Definition~\ref{def2}.


\section{Proof of Concept: Cellular Automata}
\label{sec:5}

Cellular Automata (CA) are a well-known class of discrete coupled dynamical
systems widely used in the study of complex systems and distributed
computation~\cite{mitchell1996computation}. A CA is a multi-agent system in
which every agent has a finite set of possible states, and evolves in discrete
time steps following a set of simple rules based on its own and other agents'
states. \rev{For simplicity, we focus our analysis on synchronous update
CA.\footnote{For a survey about asynchronous CA, please see
Reference~\cite{fates2013guided}.}}

CA are a natural candidate for our measures, since they have been often used in
other studies of self-organisation \cite{wolfram2002new}, some of them are
capable of universal computation~\cite{wolfram1984universality}, and they
provide a rich testbed for theories of distributed computation and 
collective behaviour in complex systems~\cite{lizier2010local}.

\subsection{Method Description}

Our analysis focuses on Elementary Cellular Automata (ECA), which constitute a
particular subclass of CA. In ECA, agents (or \emph{cells}) are arranged in a
one-dimensional cyclic array (or \emph{tape}). The state of each cell at a
given time step has two possible states, 0 or 1, and is a boolean function of
the state of itself and its immediate neighbours at the previous time step. The
same boolean function dictates the time evolution of all agents, inducing a
spatial translation symmetry. Hence, each of the 256 different boolean
functions of three binary inputs induces a different \emph{evolution rule}.
Rules are then enumerated from 0 to 255 and each ECA, irrespective of its
number of agents, can be classified by its rule. {Moreover, each rule has an
equivalent class of rules, given by the rules obtained by reflection
(exchanging right and left) and inversion (exchanging zeros and ones). Keep in
mind that all the statistical results discussed in this section are equally
valid for all the members of the corresponding equivalence class.} For a more
detailed description of ECA and their numbering system, see
Reference~\cite{wolfram2002new}.

In our simulations, we followed the protocol outlined in
Section~\ref{sec:method} over arrays of $N$ cells that followed one ECA rule.
We initialised one copy of the ECA in each of the $2^N$ possible initial
conditions and numerically computed the temporal evolution of each one of them.
As is standard in the ECA literature, the automata were simulated under
periodic boundary conditions. {The probability distribution at time $t$,
$\mu_t$, was calculated after the system reached a \textit{pseudo-stationary
regime}, which plays the role of a non-equilibrium
steady-state~\cite{esposito2010three,tome2012entropy}. These calculations were
performed using methods outlined in Appendix~\ref{app:ECA_comp}, which allowed us to
consider arrays up to size $N=17$.}

Our analysis of the ECA included the following elements:

\begin{enumerate}[leftmargin=*,labelsep=4.9mm]

    \item[(a)]The temporal evolution of $H(\boldsymbol{X}_t)$,
    $B(\boldsymbol{X}_t)$ and $R(\boldsymbol{X}_t)$. These plots show if the
    ECA shows signs of self-organisation according to Definition~\ref{def1},
    and if the joint entropy decreases or remains constant (c.f.
    Section~\ref{sec:32}).

    \item[(b)]The interdependency between individual cells through time, as
    given by the mutual information between a single cell at time $t=0$ and all
    other cells in the same and successive times (i.e. $I(X_0^0;X_t^k)$ for
      $t\in\{0,1\dots\}$ and $k\in\{1,\dots,N\}$). This reflects the predictive
      power of the state of a cell in the initial condition over the future
      evolution of the system.\footnote{To use an analogy, one can think of the
        information content of a cell as a drop of ink that is thrown into the
      river of the temporal evolution of the system.}

    \item[(c)]The mutual information between every pair of cells for the
    pseudo-stationary distribution. Because of the spatial translation symmetry
    of ECA, it suffices to take any cell and compute its mutual information
    with each other cell.  We call this ``spatial correlation,'' as it measures
    interdependencies between cells at the same time $t$. 

      \item[(e)]The curve $\psi_L$ (c.f. Section~\ref{sec:4})  \rev{for the
      pseudo-stationary distribution, which is used to} characterise a self-organising 
      system as either redundancy- or
      synergy-dominated as per Definition~\ref{def2}. This curve can also be
      interpreted as how much of a cell can be predicted by the most
      informative group of $L$ other cells.

\end{enumerate}

\subsection{Results}

Now we present and discuss the profiles of some well-known rules, which
illustrate paradigmatic behaviour. \rev{As the behaviour of ECA is known to be
sometimes affected by the specific number of agents (see e.g.
Reference~\cite{betel2013solving}), we only discuss results that are exhibited
consistently for a range of values of $N$.} Figures show results of ECA with
$N=17$ agents, while extended versions of these results for all rules with
$N=4,\ldots,17$ agents can be found in \url{https://cellautomata.xyz}.

\subsubsection{Strong redundancy: rule 232}

\begin{figure}[H]
  \centering
  \includegraphics{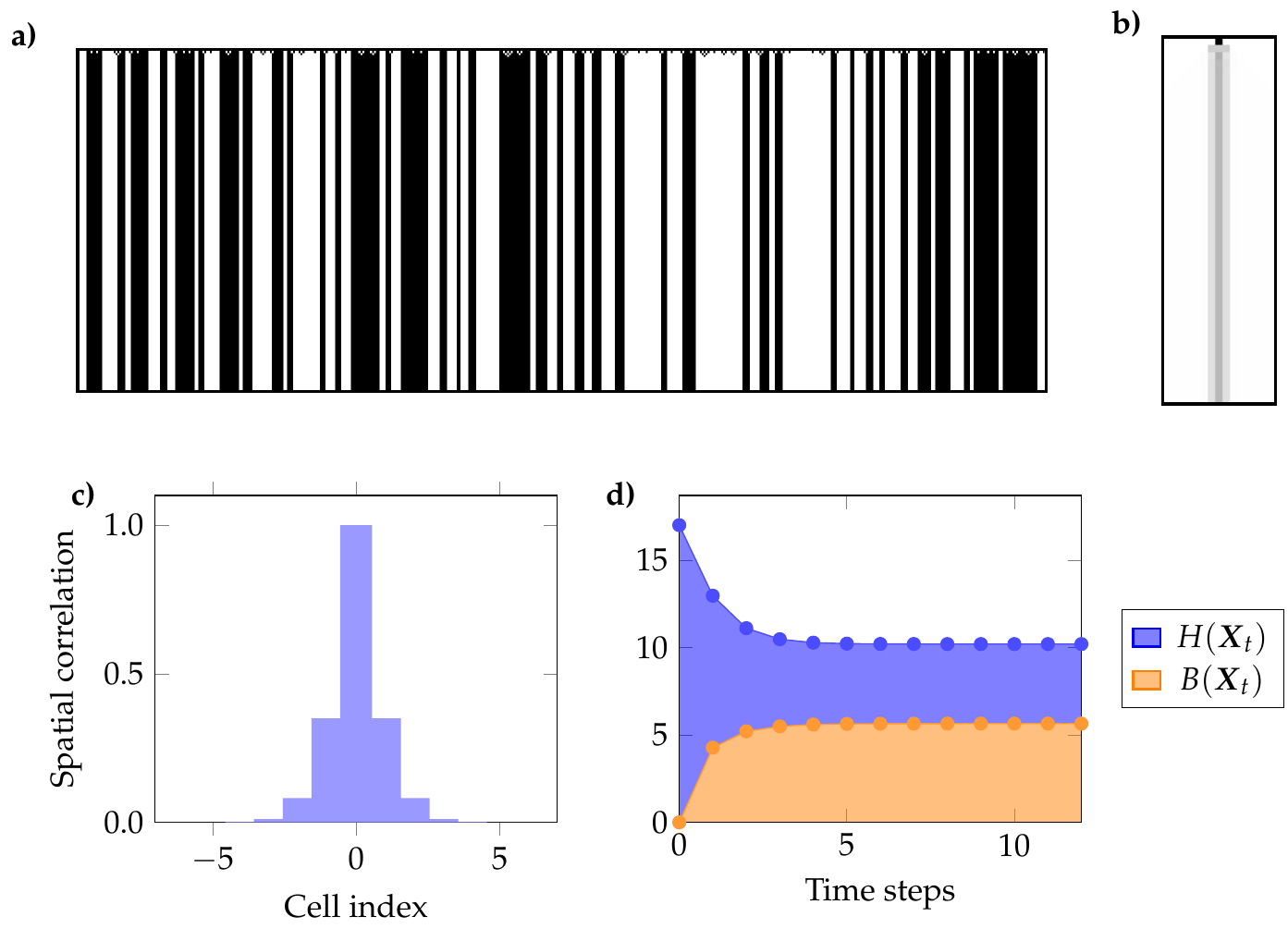}
  \caption{Combined results for rule 232. 
\textbf{(a)} Example of evolution starting from random initial conditions. \rev{Note that this example system is larger than the one used in the simulation for plots (b-d).}
\textbf{(b)} Mutual information between the initial state of a cell and the future state of the same cell and its neighbours (black is higher).
\textbf{(c)} Profile of pairwise mutual information terms between cells at the pseudo-stationary regime shows a typical exponential decay.
\textbf{(d)} Time evolution generates interaction reflected by $B(\boldsymbol{X}_t)$, which is of the same order of magnitude than $R(\boldsymbol{X}_t)=H(\boldsymbol{X}_t)-B(\boldsymbol{X}_t)$. Both $B$ and $H$ are reported in bits.}
  \label{fig:combined_plots_232}
\end{figure}

Rule 232 is commonly referred to as the majority rule, as one cell's next state
is 1 if and only if two or more of its predecessors are 1. The dynamics of this
rule when starting from a random initial condition are governed by interactions
between nearest neighbours, which are resolved after few steps into stable
configurations (Figure~\ref{fig:combined_plots_232}a). As a result of this
brief interaction, the dynamics generate binding information while decreasing the
joint entropy, as shown in Figure~\ref{fig:combined_plots_232}d.

In agreement with those observations, it is found that one cell at the initial condition 
has high predictive power over the state of itself and its nearest neighbours in
the future (Figure~\ref{fig:combined_plots_232}b). Correspondingly, the
profile of pairwise mutual information terms between cells at the pseudo-stationary 
regime shows exponentially decaying correlations as a function of cell \linebreak distance 
(Figure~\ref{fig:combined_plots_232}c).

The curve of $\psi_L$ shows a concave shape, growing strongly for the first
two (nearest) neighbours, growing slightly for the third and fourth nearest neighbours,
and remaining then essentially flat (Figure~\ref{fig:spatial_correlation_all}).
This means that remote neighbours are practically independent, which is
consistent with the pairwise correlation profile. Note that knowing all the
other cells provides an 75\% prediction over a given cell, meaning that 
there is a non-negligible amount of residual entropy.

{In summary, Rule 232 shows the signature of redundancy-dominated
self-organisation. This behaviour was found consistently in rules that evolve
towards fixed states and rules that evolve towards periodic orbits with
relatively short cycle lengths, which are known in the CA literature as Class 1
and Class 2 rules, respectively~\cite{wolfram1984universality}.}

\subsubsection{Synergistic profile: rule 30}

Rule 30 is known for generating complex geometric patterns, and has a sensitive
dependence to initial conditions~\cite{cattaneo2000investigating}. This rule,
among others, has provided key insights to understand how simple rules can
generate complex structures. For example, similar patterns can be found in the
shell of the \textit{conus textile} cone snail species. Rule 30 has also been
proposed as a stream cipher for cryptography~\cite{wolfram1985cryptography},
and has been used as a pseudo-random number generator~\cite{wolfram1986random}.

\begin{figure}[H]
  \centering
  \includegraphics{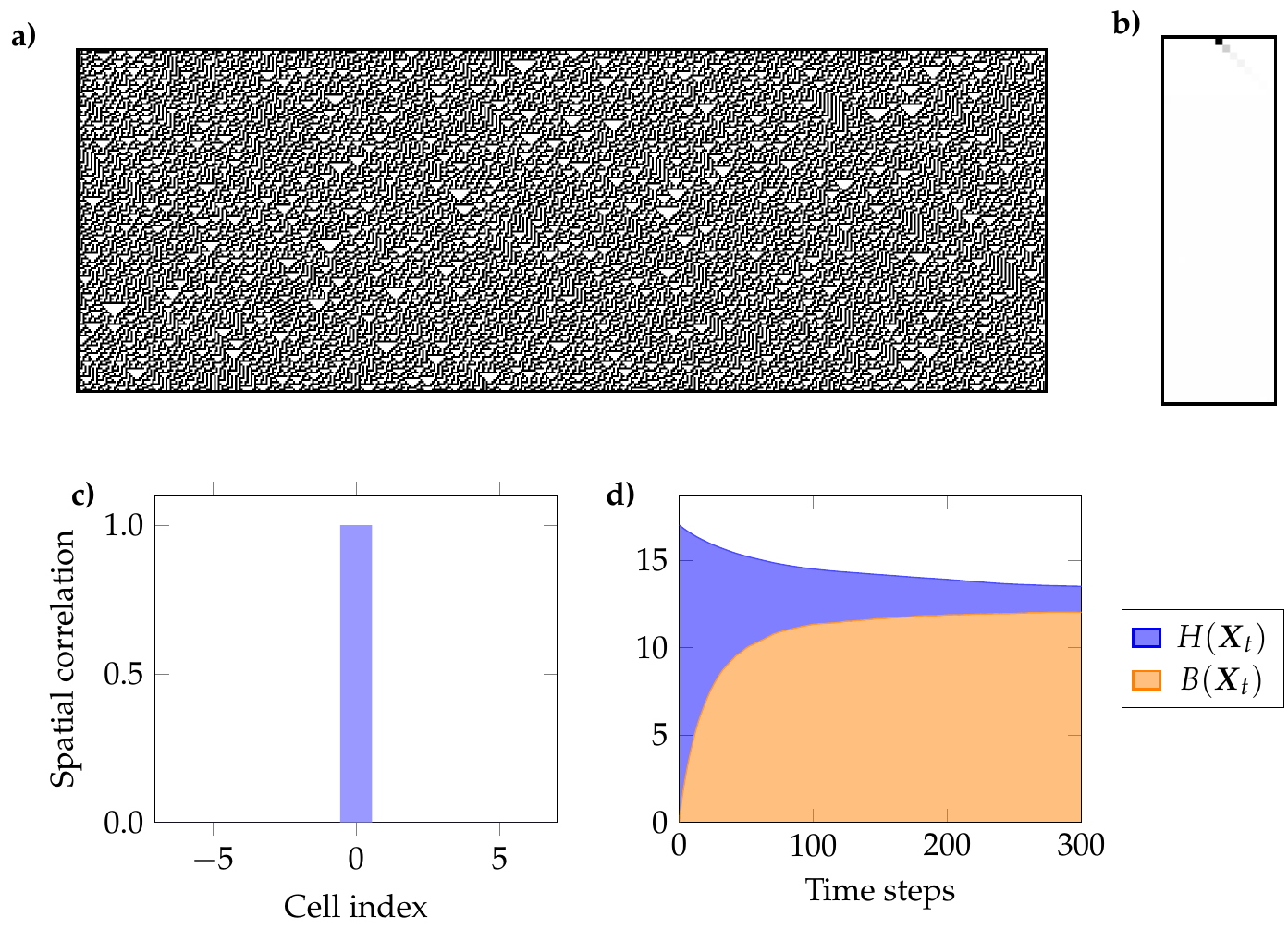}
  \caption{Combined results for rule 30.
\textbf{(a)} Example of evolution starting from random initial conditions. \rev{Note that this example system is larger than the one used in the simulation for plots (b-d).}
\textbf{(b)} Mutual information between the initial state of a cell and the future state of the same cell and its neighbours (black is higher).
\textbf{(c)} At the pseudo-stationary regime, there exists no mutual information between any pair of cells.
\textbf{(d)} Despite having no significant pairwise correlations, the dynamics generate large amounts of interdependency between the cells reflected by a high value of $B(\boldsymbol{X}_t)$. Both $B$ and $H$ are reported in bits.}
  \label{fig:combined_plots_30}
\end{figure}

Visual inspection suggests that the information processing done by this rule is
much more complex than Rule 232. In effect, Figure~\ref{fig:combined_plots_30}d
shows that this rule generates high $B(\boldsymbol{X}_t)$ through a much longer
mixing time. Intriguingly, the predictive information of a single cell seems to
disapear after very few steps (Figure~\ref{fig:combined_plots_30}b), meaning
that knowing the state of a single cell of the initial condition is not useful 
for predicting the state of any cell at later stages. Even more intriguingly, 
the pseudo-stationary regime shows that each pair of cells is practically independent 
(Figure~\ref{fig:combined_plots_30}b), in direct contrast with the high value
of $B(\boldsymbol{X}_t)$.

These aparent paradoxes are solved when one considers high-order correlations by
studying the behaviour of $\psi_L$ (Figure~\ref{fig:spatial_correlation_all}).
In effect, the convex shape of the curve shows a pronounced synergistic
structure: groups of less than 8 cells show no interdependency, but groups of
15 allow almost perfect prediction! This shows that the self-organisation
driven by Rule 30 generates high-order structures. In particular, for arrays
of 17 cells, the fact that $\sum_{j=1}^9m_j(\boldsymbol{X}_t)\approx 0$ implies
that the interdependencies are synergies of order 10 or higher.

\subsubsection{Pure synergy: rules 60 and 90}

Rule 90 consists of concatenated \texttt{xor} logic gates: the
future state of each cell correponds to the \texttt{xor} of its two precessors.
When started from a single active cell, Rule 90 generates a Sierpinsky
triangle, while when started from a random initial condition it generates
irregular triangular patterns. Rule 90 is known for having connections with
number theory, as discussed in Ref.~\cite{wolfram2002new}.

\begin{figure}[H]
  \centering
  \hspace{-1cm} \includegraphics{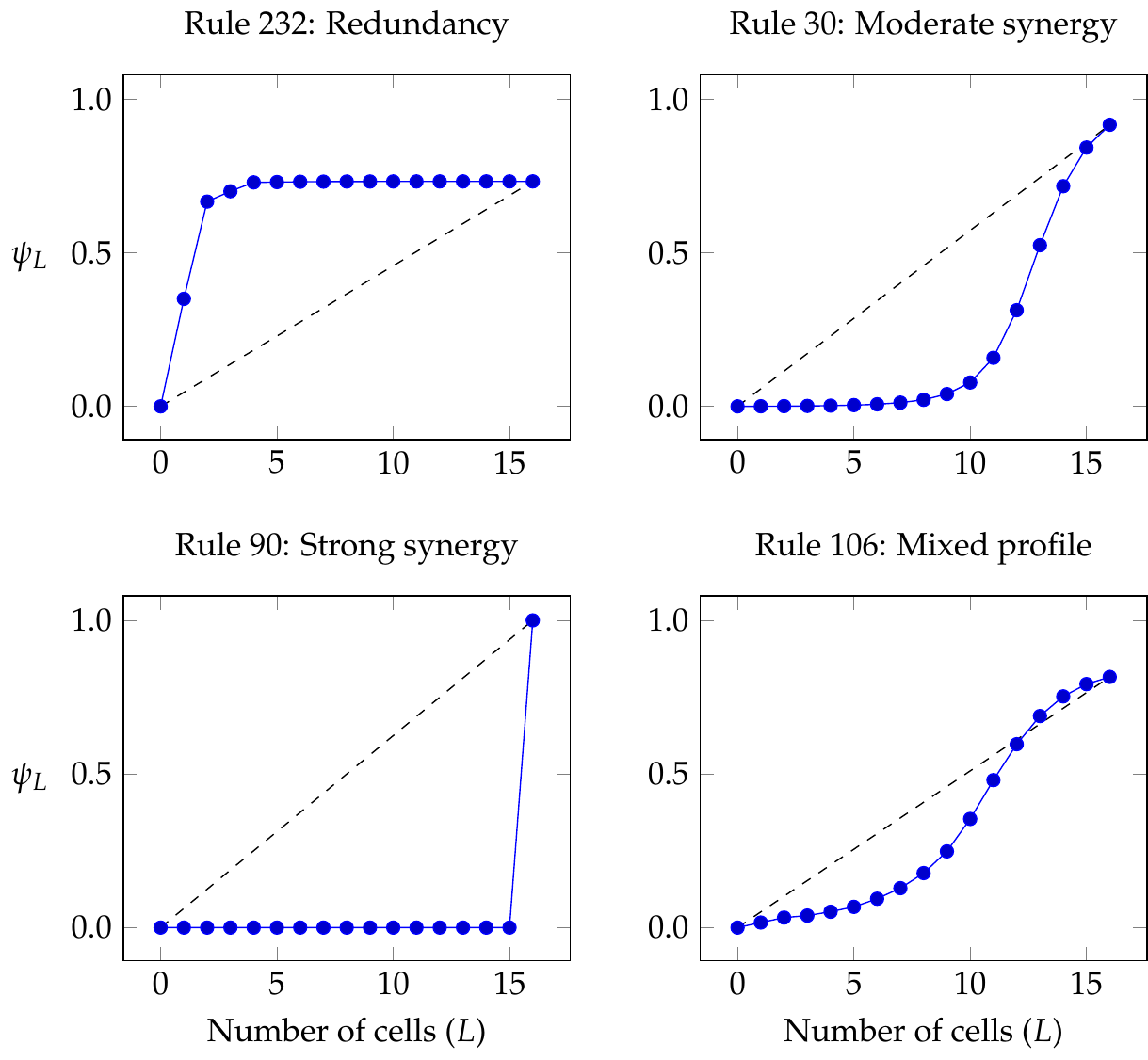}
  \caption{\rev{While the concave shape of $\psi_L$ for Rule 232 shows that
    correlations are mostly redundant, the convex shape for Rule 30 shows the
    dominance of synergies of order 10 or more. Rules 60 and 90 are the only
    rules that generate purely synergistic structure of the highest order.
    Results for Rule 106 show an inflection point where $\psi_L$ switches
  from convex to concave, suggesting the coexistence of synergistic small-scale
  and redundant large-scale structures.}}
  \label{fig:spatial_correlation_all}%
\end{figure}

{Together with Rule 60, which is also composed by concatenated \texttt{xor}s,} 
Rule 90 was found to be the most synergistic rule of all 256 ECA. In fact, for
an array of $N$ cells started with random initial conditions, after the second step 
any group of $N-1$ cells or less is statistically independent. This implies that 
$\psi_L=0$ for all $L<N-1$, and therefore $m_L(\boldsymbol{X}_t)=0$ for all
$L<N-1$ (Figure~\ref{fig:spatial_correlation_all}). However, our calculations show that
$R(\boldsymbol{X}_t)=0$ while $B(\boldsymbol{X}_t)=H(\boldsymbol{X}_t)=N-1$,
indicating that the binding information of Rules 60 and 90 corresponds exclusively
to synergy of the highest order, i.e. $B(\boldsymbol{X}_t) =
m_{N-1}(\boldsymbol{X}_t)$.

{We found that most ECA rules with attractors of length of the order of the
phase space (known as Class 3 and 4 in the CA
literature~\cite{martinez2013computation}) exhibit synergy-dominated
self-organisation. Besides rules 30, 60 and 90 (and the ones in their
equivalence classes), rules 18 and 146 have the strongest convexity in their
$\psi_L$ profiles.} Interestingly, the fact that rules 60 and 90 have been
found to have the highest synergy is consistent with the crucial role played by
\texttt{xor} gates in cryptography.\footnote{For a discussion of this
connection, see Reference~\cite{rosas2016understanding} and Section 4.2.}

\subsubsection{Coexistence of Convex and Concave Segments in $\psi_L$}
\label{sec:mixed_profile}

{
Interestingly, some rules show both convex and concave sections in
$\psi_L$. Examples of this phenomenon are rules 14, 22, 41, 54, 62, 73, 106
and 110, with the shape of $\psi_L$ being sometimes sensitive to the system
size. Rules 106 and 110, in particular, show a clear distinction between a
convex segment for small $L$ and concave segment for large $L$. When compared
with rule 106, rule 110 has its inflection point at a smaller $L$, which could
be related with the more localised structures seen in this rule.}

{
Based on these results, we hypothesise that a combination of synergy and
redundancy within a single system could provide a richer, or more ``complex''
structure. However, further investigation in larger systems would be necessary
to confirm that the inflection point is actually an intrinsic property of the
rule -- and not a finite-size effect.}

\section{Discussion}
\label{sec:6}
 
This paper presents an information-theoretic framework to study
self-organisation in multi-agent systems, which explores how statistical
structures are spontaneously generated by the evolution of coupled dynamical
systems. To guarantee the absence of centralised control guiding the process,
we restrict ourselves to autonomous systems where each agent can interact
directly with only a small number of other agents. To isolate structures that
are purely created by the system's dynamics, we consider the evolution of
agents that are initially maximally random and independent.

A fundamental insight behind our framework is the fact that
\textit{deterministic dynamical systems are able to create correlations by
destroying information}. In effect, we saw that while the temporal evolution of
many dynamical systems reduce their joint Shannon entropy, this condition can
be the consequence of two qualitatively opposite scenarios: in one case
interdependency is created while the stochasticity of each agents is preserved;
and in the other mere information dissipation occurs (each agent becomes less
random while remaining independent of each other). Following this line of
thought, and diverging from the standard literature, we propose to attribute
self-organisation to processes where the strength of interdependencies increases with
time. In this work we use the binding information as metric of global
interdependency strength.

As a second step, we propose a multi-layered description of the attained
organisation based on synergies and redundancies of various orders. The key
idea is to decompose the information stored in the system, as quantified by the
joint Shannon entropy, considering two principles: extension (how many agents
are linked), and sharing mode (how many agents need to be measured in order to
obtain predictive power). The information sharing mode of order 1 corresponds
to redundancy, which takes place when by measuring only one agent one can
(partially) predict the state of a number of other agents. Synergy takes place
when such predictive power is accessible only when measuring two or more
agents simultaneously. We proposed these decompositions as formal structures, without
providing an explicit way to compute the values of their components for
arbitrary probability distributions. Nevertheless, upper and lower bounds for
these components are provided, which in some cases can allow a complete
determination of the decomposition.

Using the proposed framework, this work is the
first -- to the best of our knowledge -- to demonstrate cases of high-order
statistical synergy in relatively large systems. In particular, we showed that the ECA that
corresponds to rule 90 generates maximal synergy, which, according
Reference~\cite{rosas2016understanding} and Section 4.2, could enable the development
of interesting cryptographic applications. {Moreover, our results suggest
that some rules can exhibit a coexistence of redudant and synergistic
structures at different scales. However, more work is needed in order to
confirm this hypothesis and explore its implications.}

{Let us remark that our framework does not intend to compare diverse
systems on a unidimensional ranking of organisational richness. Accodingly, it
would not be correct to claim that rule 90 attains a richer organisation than
other ECA. Our framework uses increments in the binding information to detect
self-organisation, and then applies an multi-dimensional information
decomposition to provide qualitative insight of the result of this process. As
a result, different types of structures (e.g. redundant, synergistic or mixed)
are acknowledged in their diversity, without trying to collapse their
properties into a single number.}

An interesting extension of this work would be to use some of the recently
proposed measures of synergy (see e.g.
\cite{ince2017measuring,jamesdit,makkeh2018broja,finn2018pointwise}) to build
exact formulas for the proposed decompositions. {This would allow a more
precise characterisation of the strength of each information sharing mode.}
However, this could prove to be challenging, as most of these metrics are
designed for systems of three variables with their extensions to larger systems
not being straightforward.

{
Another natural extension would be to apply the presented framework to study
continuous coupled dynamical systems, and also their stochastic counterparts
(e.g. stochastic differential equations). Interestingly, while the entropy of
continous systems can be negative, the binding information is still a
non-negative quantity and hence its decomposition can be carried out directly
using the framework proposed in Section~\ref{sec:34}. Moreover, all the
presented results and methods are valid for systems with random dynamics, with
the sole exception of the fact that the joint entropy can increase (in contrast
to what was discussed in Section~\ref{sec:32}). The main challenge for this
would be to develop faithful estimators of the corresponding densities for
cases where analytical expressions are not available. This task could, for
example, be approached by using well-established methods of Bayesian
inference~\cite{gelman2013bayesian} and density
estimation~\cite{Bishop:2006:PRM:1162264}.}

{
To study the structure of a particular attractor in a non-ergodic system, one
could focus the analysis on the corresponding natural invariant measure (c.f.
\cite{young2002srb}) instead of studying the evolution from the uniform
distribution. It would be of interest to explore if well-known chaotic
attractors can be explained in terms of the synergies and redundancies they
induce in the corresponding coordinates, which could provide a new link between
chaos theory and multivariate information theory. These developments could
allow to study real-world phenomena, e.g. sensorimotor control loops
\cite{nurzaman2015goal,sandor2015sensorimotor}. Also, this development could 
enable a bridge between the
ideas presented in this paper and the extensive literature of self-organising
coupled oscillators (see e.g. \cite{pikovsky2003synchronization,kuramoto2012chemical}).}

Finally, it is worth emphasising that the statistical character of our proposed
framework makes it orthogonal to some well-established self-organisation
principles, such as the \textit{enslaving principle} for multi-scale
systems~\cite{haken1977synergetics} or the \textit{free energy principle} for
autopoetic organisms~\cite{friston2010free}. As a matter of fact, it remains to
be explored to what extent those principles can be enriched by including
multi-layered decompositions in terms of redundancies and synergies. Also,
please note that the presented approach to self-organisation is restricted to
structures that are generated within known possibilities, which is related to
the idea of ``weak emergence''~\cite{chalmers2006strong}. An attractive
extension would be to include phenomena related to ``strong emergence,'' i.e.
processes in which evolution can affect the state space itself, generating
entirely new configurations for the system to explore. An attractive way of
attempting this extension could be to combine the presented framework with the
notion of super-exponentially growing phase spaces presented in
Reference~\cite{jensen2018statistical}.


\vspace{6pt}
\authorcontributions{All authors participated in the development of the concepts, wrote and revised the manuscript. The numerical evaluations over ECA were carried out by M.U. All authors have read and approved the final manuscript.}

\funding{FR was supported by the European Union’s H2020 research and innovation programme, under the Marie Sk\l{}odowska-Curie grant agreement No. 702981.}

\acknowledgments{The authors thank Carlos Gershenson, Michael Lachmann, Robin Ince and Ryan James for helpful discussions. The authors also acknowledge useful suggestions from Karl Friston, Tiago Pereira and an anonymous reviewer, which greatly improved the paper.}

\conflictsofinterest{The authors declare no conflict of interest. The funding sponsors had no role in the design of the study; in the collection, analyses, or interpretation of data; in the writing of the manuscript, and in the decision to publish the results.} 



\appendixtitles{yes} 
\appendixsections{multiple} 
\appendix

\section{Alternative approaches to formalising global structure}
\label{app:alternative_structure}

\subsection*{Structure as Geometrical Properties of Attractors}

A natural way to attempt a formalisation of structure is by relating it to the
notion of \textit{attractor} from the dynamical systems literature. To define
what an attractor is, let us note that the set
$\{\boldsymbol{x}_s | \boldsymbol{x}_s = \phi_{s}(\boldsymbol{x_0})\}$
corresponds to the \textit{trajectory} that originates from the initial
condition $\boldsymbol{x}_0 \in \Omega$. A set $B\subset \Omega$ is stable if
it contains the trajectories of all its elements, i.e. for all
$\boldsymbol{y}_0 \in B$ and $t \in \mathbb{N}$ we have that
$\boldsymbol{y}_t=\phi_t(\boldsymbol{y}_0)\in B$. An attractor is a set $A$ that
is stable and has no stable proper (non-empty) subsets. Common attractors are
fixed points, limit cycles (i.e. periodic trajectories) and strange
attractors~\cite{strogatz2018nonlinear}.

Since the early efforts of
Ashby~\cite{ashby1947principles,ashby1962principles}, it has been noted that
self-organisation is a consequence of the tendency of dynamical systems to evolve
towards attractors. The system, hence, becomes more ``selective'' as time passes. Following this
rationale, one can argue that the distinctive properties of these attracting
configurations are the ones that emerge within the time-evolution.

Could one relate the attractor's geometrical structure to properties of
structure and organisation? An attractive fact is that ``interesting'' dynamics
are usually associated with non-linear equations, which in turn generate
strange attractors with exotic geometric properties; while on the other hand
the attractors of linear dynamics have uninteresting geometrical structure.
Following this line of thought, one could attempt to establish relationships
between the geometrical structure of attractors (e.g. in terms of fractal
structure) and properties of the organisation attained by the agents. Although
this is plausible, the route to develop such endeavour is not
straightforward.

\subsection*{Structure as Pattern Complexity}
\label{sec:Kc}

Interesting approaches for formalising the concept of structure or pattern can
be found in the computer science and signal processing literature. 
One such approach is to relate pattern strength with the
Kolmogorov complexity (KC)~\cite{kolmogorov1965three}, which is the length of
the shortest computer program that is able to generate the pattern as output.
In this way, pattern stength is inversely proportional to the value of the
corresponding KC: very structured configurations can be generated by short
programs, while random configurations with no structure can only be the output
of a program of the same length of the sequence itself.

Using the KC to measure pattern strength is attractive due to its
intuitiveness, and because its quantitative nature can allow comparisons
between heterogenous structures~\cite{li2008p}. Unfortunately, the KC has been
proven to be not computable (this imposibility being related to G\"{o}del's
incompletness theorem~\cite{chaitin1990information}), which hinders its
practical value.\footnote{Please note that practical estimation of the KC can be
attempted via upper bounds, which can be calculated using lossless compression
algorithms~\cite{li2008p}.} Another weakness of this approach is that the KC
does not have, to the best of our knowledge, properties about the relationship
between the complexity of a system and the complexity of it parts, nor
properties of how the KC evolves in time under diverse dynamical conditions.
These two limitations are overcomed by adopting an information-theoretic
framework, as we do in the main body of the paper.

\section{From a Dynamical System to a Stochastic Process}
\label{ap:technical}
In order to consider probabilities defined over a metric phase space $\Omega$, one
needs to introduce a collection of ``events,'' denoted by $\mathcal{B}$, 
\rev{which correspond to measurable subsets of $\Omega$. It is natural to ask 
this collection to be a $\sigma$-field, so that if $B_1,B_2 \in \mathcal{B}$ then 
$B_1\cup B_2\in \mathcal{B}$ and $B_1\cap B_2\in \mathcal{B}$ are 
guaranteed~\cite{Loeve1978}. A probability measure $\mu$ is a function 
$\mu: \mathcal{B} \mapsto [0,\infty)$
such that $\mu(\Omega) = 1$, which satisfies the relationship 
$\mu\left( \cup_{j=1}^\infty B_j \right) =
\sum_{j=1}^\infty \mu(B_j)$ if $B_j \in \mathcal{B}$ for all $j \in \mathbb{N}$ 
and $\cap_{j=1}^\infty B_j=\varnothing$.} When considering a map 
$\phi_t$ over the phase space, it is natural 
to require $\phi_t$ and $\mathcal{B}$ to match together appropiately, i.e. for all
$B\in\mathcal{B}$ then $\phi_t^{-1}(B) = \{\boldsymbol{x} \in \Omega|
\phi_t(\boldsymbol{x}) \in B\} \in \mathcal{B}$. In that way, one can guarantee
the consistency of the definition given in \eqref{eq:flow_measure}.

Given a probability distribution $\mu_0$, any measurable function $Y: \Omega
\mapsto \mathbb{R}$ can be considered to be a random variable with statistics
defined as
\begin{equation}
\mathbb{P}\{ Y \in I \} \coloneqq \mu\big( Y^{-1}( I)\big) 
\enspace.
\end{equation}
Above, $Y^{-1}(I) = \{ \boldsymbol{x}\in \Omega| Y(\boldsymbol{x}) \in I \}$
and $I \subset \mathbb{R}$. Similary, the multivariate stochastic process
$\boldsymbol{X}_t=(X_t^1,\dots,X_t^n)$ induced by the map $\phi_t$ is
defined by the joint statistics are given \rev{by
\begin{equation}
\mathbb{P}\{ X^{i_1}_{t_0} \in I_1,\dots, X^{i_m}_{t_m} \in I_m\} \coloneqq \mu_0\left( \cap_{j=1}^m \left\{ \boldsymbol{x}\in\Omega \Big|  \phi^{j}_{t_j}(\boldsymbol{x}) \in I_j \right\} \right)
\enspace,
\end{equation}
where $i_j \in \{1,2,\dots,n\}$ are a collection of indices, $t_1,\dots,t_m \in
T$ is a collection of time points, $I_j \subset \mathbb{R}$ for
$j=1,\dots,m$, and $\phi_t^j$ is the $j$-th coordinate of the map at time $t$ as defined 
in Section~\ref{sec:333}}. For discrete phase spaces, the joint probability distribution of 
$\boldsymbol{X}_t$ is given by $p_{\boldsymbol{X}_t}(\boldsymbol{x}) = 
\mathbb{P}\{ \boldsymbol{X}_t=\boldsymbol{x}\}$ for $\boldsymbol{x}\in\Omega$.

\section{Information and Entropy}
\label{ap:entropy}

The entropy is a functional over the probability distribution that describes the state of knowledge that an observer has 
with respect to a given system of interest~\cite{jaynes2003probability}. In this context, uncertainty in the system corresponds to information that 
can be potentially extracted by performing adequate measurements.

Following this line of thought, the amout of information needed to specify a
single configuration within $|\Omega|$ possibilities is $\log |\Omega|$, where
the base of the
logarithm can be choosen according to the preferred units for counting
information (bits, nats, or others). If a system with a phase space of
cardinality $|\Omega|$ at time $t$ follows a statistical distribution
$p_{\boldsymbol{X}_t}$, then this information gets divided as
follows~\cite{rosas2016understanding}:
\begin{equation}
\log |\Omega| = H(\boldsymbol{X}_t) + \mathcal{N}(\boldsymbol{X}_t),
\end{equation}
where $ H(\boldsymbol{X}_t) \coloneqq - \mathbb{E}\left\{ \log
p_{\boldsymbol{X}_t}(\boldsymbol{X}_t) \right\}$ is the joint Shanon entropy of
the system, and $\mathcal{N}(\boldsymbol{X}_t) \coloneqq \log |\Omega| -
H(\boldsymbol{X}_t)$ is the ``negentropy''. After an observer comes to know the
statistics of the system, as encoded by $p_{\boldsymbol{X}_t}$, the average
amount of information needed to specify a particular configuration decreases
from $\log |\Omega|$ to $H(\boldsymbol{X}_t)$; therefore, the negentropy
corresponds to the bits that are disclosed by the knowledge of the statistics.
In contrast, the Shannon entropy measures the information that is not disclosed
by the statistics, which can only be obtained when the configuration of the
system is actually measured.

\section{Simulation Details}
\label{app:ECA_comp}

This appendix describes the procedure for calculating the evolution of probability distributions over ECA (c.f. Section~\ref{sec:6}).
The possible states of an ECA with $N$ cells were encoded as a binary numbers, and hence the phase space corresponds to $\Omega=\{0,\dots,2^N-1\}$. Probability distributions over the phase space were stored as an array $L_\mu = (\mu(0),\dots,\mu(2^N-1))$, where $\mu(k)\geq 0$ for all $k\in\{0,\dots,2^N-1\}$ and $\sum_{k=0}^{2^N-1}\mu(k)=1$.

We considered trajectories over $\Omega$, which correspond to sequences of binary numbers $(s_1,s_2,\dots)$ such that $s_{k+1} = \phi(s_k)$ with $\phi: \Omega\to\Omega$ being the function that encondes the ECA rule. Our interest was to find the step at which $\phi(\cdot)$ brings the trajectory back to a state that has been already visited before. Note that all trajectories end up in a periodic attractor, this being a consequence of the finiteness of $\Omega$. From a trajectory starting at $s\in\Omega$, we store the pair $(p_s, a_s)$ with $p_s$ being the length of the trajectory until reaching a state in the periodic attractor, and $a_s$ being the legth of the periodic attractor (i.e. the number of states between the first and the second appearance of a repeated state). The interest of these numbers lays in the fact that
\begin{equation}
    \phi_t(s) = \phi_{K(t,s)}(s) 
\qquad\forall t\in\{p_s,p_s+1,\dots\}
\enspace,
\end{equation}
where $K(t,s) \coloneqq p_s+(t-p_s)\mod a_s$. \rev{Above, $\phi_t(s) = \phi \circ \dots \circ \phi(s)$ is the $t$-th composition of $\phi$ with itself.}

The ECA is said to have reached a \textit{pseudo-stationary regime} when it has been run for a number of steps $t_\text{s}$ such that, for any initial distribution $\mu_0$, a trajectory of distributions $(\mu_0,\mu_1,\dots,\mu_{t_\text{s}})$ obtained by time evolution would reach a distribution that has already been visited before.\footnote{As mentioned in Section~\ref{sec:31}, the set of all probability distributions $\mu$ over $\Omega$ together with their dynamics form a new dynamical system, which also has periodic attractors. From this point of view, $t_\text{s}$ is the smallest integer such that all trajectories of distributions reach their periodic attractor.} The minimal number of steps needed to a reach pseudo-stationary regime, denoted by $t_0$, can be calculated as
\begin{equation}
t_0=LCM\big( \{a_s\}_{s\in\Omega} \big) + \max_{s\in \Omega} p_s\ ,
\end{equation}
where $LCM$ stands for the \textit{least common multiple}. Above, the last term ensures that each state entered their periodic attractor, and the former is the smallest number of steps that guarantees a simultaneous full cycle of all the attractors. For the considered ECA with $N=17$ cells, the largest values found where $t_0 \approx 10^{14}$.

In order to be able to study the statistics of ECA under pseudo-stationary regimes, we developed an efficient way to compute the evolution of a given initial distribution $\mu$ for very large number of steps. Let us represent the initial distribution $\mu$ by the array $L_{\mu}$, and the resulting distribution after $t_0$ steps as $\mu'$ with its corresponding vectorial representation $L_{\mu'}$. Our key idea is to compute the trajectory from each $s\in\Omega$ only for $K(t_0,s)$ steps -- as additional multiples of $a_s$ correspond to mere cycles over its periodic attractor. The general procedure for computing $\mu'$ using this idea goes as follows:
\begin{itemize}[leftmargin=*,labelsep=4.9mm]
\item[1.] Initialize the components of $L_{\mu'}$ with zeros.
\item[2.] For each $s\in\Omega$: compute $s'=\phi_{K(t_0,s)}(s)$ and then add $\mu(s)$ to $\mu'(s')$ (i.e. add $L_\mu[s]$ to $L_{\mu'}[s']$). 
\end{itemize}
This technique resulted to be very efficient, as the largest values found over all ECA rules for $N=17$ were $\max_{s\in \Omega} p_s= 1776$ and $\max_{s\in \Omega} a_s=78821$.


\reftitle{References}






\end{document}